\documentclass{elsart}
\usepackage{amssymb}
\usepackage{amsmath}
\usepackage{bm}
\expandafter\ifx\csname package@font\endcsname\relax\else
 \expandafter\expandafter
 \expandafter\usepackage
 \expandafter\expandafter
 \expandafter{\csname package@font\endcsname}%
\fi
\begin{document}
\begin{frontmatter}
\title{On the Speed of Gravity and Relativistic $v/c$ Corrections to the 
Shapiro Time Delay}
\author[sk]{Sergei M. Kopeikin\corauthref{cor1}}\ead{kopeikins@missouri.edu}\and\author[ef]{Edward B. Fomalont}\ead{efomalon@nrao.edu}
\address[sk]{Deptartment of Physics \& Astronomy, University of
Missouri-Columbia, Columbia, MO 65211, USA}
\address[ef]{National Radio Astronomy Observatory, Charlottesville, VA 
22903, USA}
\corauth[cor1]{Corresponding author}
\begin{abstract}
Recent papers by Samuel \cite{samuel,sam2} declared that the linearized post-Newtonian $v/c$ effects are too small to have been measured in the
recent experiment involving Jupiter and quasar J0842+1845 \cite{kajl,pr1,fk-apj} that was used to measure the ultimate speed of gravity defined as a fundamental constant entering in front of each time derivative of the metric tensor in the Einstein gravity field equations. 
We describe our Lorentz-invariant formulation of the Jovian
deflection experiment and confirm that $v/c$ effects are do observed,
as contrasted to the erroneous claim by Samuel, and that they vanish if and only if the speed of gravity is infinite.
\end{abstract}

\maketitle\end{frontmatter}
Gravitational light-ray deflection experiments testing time-dependent components of the gravitational field are important for deeper understanding of the causal (null-cone) structure of the space-time manifold and the Einstein theory of general relativity. General relativity predicts that a light particle (photon) is deflected by the gravitational field of a moving body from its retarded, with respect to observer, position \cite{ksge,ks,km}. This is a direct physical consequence of the relativistic nature of the gravitational field which is expressed mathematically in terms of the retarded Lienard-Wiechert solution of the linearized Einstein equations \cite{LL}. We proposed relativistic VLBI experiment \cite{kajl} to measure this effect of retardation of gravity by the field of moving Jupiter via observation of light bending from a quasar and successfully completed this experiment in September 2002. General relativistic experimental model of VLBI data processing algorithm is described in \cite{pr1} and the results of the main experiment of September 2002 are published in the Astrophysical Journal \cite{fk-apj}.

Recently Samuel \cite{samuel,sam2} made an attempt to review the results of the experiment by making use of a linearized Lorentz transformation of the static Shapiro time delay. This approach is physically meaningful but correct conceptual analysis and interpretation of the experiment requires matching of the first, $v/c$, and second $v^2/c^2$ order effects in the relativistic theory of the Shapiro time delay which has neither been conceived nor completed in \cite{samuel,sam2}. This led Samuel \cite{samuel,sam2} to the confusion of the Lorentz transformation of the gravitational field and that of light used for observations. The same conceptual flaw is contained in papers by Asada \cite{asada} and Will \cite{wil} who were not able to separate a general-relativistic effect of the aberration of gravity from the special-relativistic effect of the aberration of light \cite{kop4}. Moreover, Samuel \cite{samuel,sam2} has arrived to erroneous conclusion that the post-Newtonian effect of order of $v/c$ beyond the static part of the Shapiro effect can not be observed at all. In the present {\it Letter} we outline the basic equations of the {\it exact} Lorentz-invariant theory of the gravitational time delay of light and show mistakes in Samuel's interpretation \cite{samuel,sam2} of the gravitational light-ray deflection experiments.

For Jupiter, the perturbation of the metric tensor, $h^{\mu\nu}$, in harmonic
coordinates $x^\alpha=(ct, {\bm x})$ is obtained after solving the linearized Einstein equations in terms of the retarded Lienard-Wiechert gravitational potentials \cite{LL}. It yields \cite{kajl,fk-apj,ks} 
\begin{equation}\label{1} 
h^{\mu\nu}=-\frac{2GM_J}{c^4}\frac{2u^\mu u^\nu+\eta^{\mu\nu}}{r_\alpha 
u^\alpha}\;.
\end{equation}
Here \footnote{Greek indices run from 0 to 3. Bold letters denote spatial vectors. Repeated indices mean 
the Einstein summation rule. Euclidean dot and cross products of two vectors are 
denoted as ${\bm a}\cdot{\bm b}$ and ${\bm a}\times{\bm b}$ respectively. Partial derivative  with respect to coordinate $x^\alpha$ is denoted as $\partial_\alpha\equiv\partial/\partial x^\alpha$.} $\eta^{\mu\nu}$ is the Minkowski metric, $M_J$ is the mass of
Jupiter, $u^\alpha=\gamma(s)(c, {\bm v}_J(s))$ is its four-velocity,
with ${\bm v}_J(s)=d{\bm x}(s)/ds$ and
$\gamma(s)=(1-v_J^2(s)/c^2)^{-1/2}$.  The distance
$r^\alpha=x^\alpha-x^\alpha_J(s)$, and Jupiter's worldline
$x^\alpha_J(s)=(cs, {\bm x}_J(s))$ are functions of the retarded time $s$,
determined as a solution of null-cone equation for gravity 
\begin{equation}
\label{gnc}
\eta_{\mu\nu}r^\mu r^\nu=0\;.
\end{equation}
Substituting $r^\mu(s)$ and solving (\ref{gnc}) in the causality region, $t>s$, show that the retarded time $s$ obeys a (non-linear) equation 
\begin{equation}
\label{2}
s=t-\frac{1}{c}|{\bm x}-{\bm x}_J(s)|\;,
\end{equation}
which describes the null-cone characteristics of Jupiter's gravity field with $c$ being understood as the ultimate speed of gravity. Retarded time $s$ appears as an argument in coordinates of Jupiter and reflects the obvious fact that gravity field is causal and propagates with finite speed $c$ as predicted by Einstein's theory of general relativity \cite{LL}. Despite that this derivation is well-known and given in the advanced course on general relativity \cite{LL} some researchers \cite{asada,pask} are still confused about the nature of the retarded-time equation (\ref{2}) in the gravitational Lienard-Wiechert potentials (\ref{1}) and misinterpret it as a classic R\"omer time delay of light propagating from a massive body (Jupiter) to the field point.

VLBI measures the phase $\phi$ of the radio wave front coming from a quasar.  Diffferential equation for the phase is determined by the first integral of the light-ray geodesics that is known as the eikonal equation  \cite{LL} 
\begin{equation}
\label{eik}
g^{\mu\nu}\partial_\mu\phi\partial_\nu\phi=0\;,
\end{equation}
where the metric tensor
$g^{\mu\nu}=\eta^{\mu\nu}-h^{\mu\nu}$. Provided that one neglects orbital acceleration of Jupiter, 
solution of equation (\ref{eik}) is given by \cite{found,kwtn}
\begin{equation}
\label{3}
\phi=\phi_0+\frac{\nu}{c}\left[k_\alpha x^\alpha+\frac{2GM_J}{c^2}\left(k_\alpha 
u^\alpha\right)\ln\left(k_\alpha r^\alpha\right)\right]\;,
\end{equation}
where $\phi_0$ is a constant, $\nu$ is the radio frequency, $k^\alpha=(1,{\bm k})$ is a
null vector of the radio wave, and the unit vector ${\bm k}$ is directed from the quasar to VLBI station. Notice that the argument of the logarithm in equation (\ref{3}) depends on the retarded time $s$ defined by the gravity null-cone equation (\ref{2}). It makes evident that the phase of electromagnetic wave passing through time-dependent gravitational field of moving Jupiter is affected by this field not instantaneously but with a delay being required for gravity to propagate from the retarded position ${\bm x}_J(s)$ of Jupiter to the position ${\bm x}$ of a radio photon. Measuring this `retardation-of-gravity' effect in the light-ray deflection experiments like that we had conducted in September 2002 \cite{kajl,fk-apj}, allows one to corroborate that gravity propagates on a null cone with finite speed \cite{kajl,fk-apj,found}. 

Equation (\ref{3}) is Lorentz-invariant and was obtained by making use of the Lienard-Wiechert gravitational potentials (\ref{1}). It can be also derived by solving the eikonal equation (\ref{eik}) in a co-moving frame of Jupiter with subsequent Lorentz transformation of both gravitational and electromagnetic field from this frame to the barycentric frame of the solar system \cite{kop4,klio,kop5}. This point is subtle and was not properly understood by Samuel \cite{samuel,sam2} and some other researchers \cite{asada,wil,carlip,faber} who erroneously believe that the linearized (that is, $v/c$) approximation of the Lorentz transformation affects exclusively the direction of propagation of radio wave used for observation of the gravitational light-ray deflection but leaves the gravitational field of Jupiter in the barycentric frame of the solar system static (see also \cite{wsg}). The message of this {\it Letter} is that in the barycentric frame of the solar system the gravitational field is characterized in the neigborhood of Jupiter not only by the components of the metric tensor $g_{\alpha\beta}$ but its space and time derivatives (the Christoffel symbols and the curvature tensor) as well. The eikonal equation (\ref{eik}) is the first integral of the light geodesic equations \cite{LL}. This integral can be present in the form given by equation (\ref{eik}) if and only if the speed of gravity and the speed of light are equal or, in other words, the time derivatives of the metric tensor in the Einstein equations are normalized to the same fundamental speed as the time derivatives of electromagentic field in the Maxwell equations. Lorentz transformation from Jupiter's frame to the barycentric frame of the solar system induces non-zero values of the time derivatives of the metric tensor if the speed of gravity is not infinite \footnote{Notice that the speed of physical light does not play any role in the Lorentz transformation of the gravitational field.}. In general relativity the effect of these first time derivatives of the metric tensor cancel with the aberration of light when one integrates equation of light geodesics to obtain the eikonal equation (\ref{eik}) which leaves the retardation of gravity effect present in Jupiter's coordinates ${\bm x}_J(s)$ in equation (\ref{3}). Therefore, the retardation of gravity effect appears in the linearized order of $v/c$ in the post-Newtonian expansion of the retarded coordinate ${\bm x}(s)$ of Jupiter around the time of observation $t$ \cite{kajl} as we demonstrate in the next paragraph. It can be easily confused with the aberration of light if the effect of the first time derivatives of the metric tensor is not pursued in derivation of equation (\ref{3}) for the electromagnetic phase. The reason for the `aberration-of-light' misinterpretation of the Jovian light-ray deflection experiment given in papers \cite{samuel,sam2,asada,wil,pask,carlip,faber} is that the Nordtvedt-Will PPN formalism \cite{will} has a faulty assumption leading to the replacement of the speed-of-gravity parameter $c_g$ in front of the first time derivatives of the metric tensor with the physical speed of light \cite{kop4,kop5}. This makes the PPN gravitational force acting on test particles, be dependent upon the speed of electromagnetic waves. Thus, the PPN supposes that the gravitational force is tightly connected with Maxwell's theory which is obvously a wrong physical postulate leading to a conceptual failure of the Nordtvedt-Will PPN formalism in case of time-dependent gravitational fields.  

Let us now consider the post-Newtonian expansion of the eikonal (\ref{3}). Toward this end we approximate 
Eq.~(\ref{3}) by introducing two angles $\Theta$
and $\theta$ between vector ${\bm k}$ and unit
vectors ${\bm p}={\bm R}/R$ and ${\bm l}={\bm r}/r$, where vectors ${\bm
R}={\bm x}-{\bm x}_J(t)$ and ${\bm r}={\bm x}-{\bm
x}_J(s)$ connects 
the point of observation ${\bm x}\equiv{\bm x}(t)$ with the present, ${\bm x}_J(t)$, and retarded, ${\bm x}_J(s)$, positions of Jupiter, respectively. By definition $\cos\Theta={\bm k}\cdot{\bm p}$ and
$\cos\theta={\bm k}\cdot{\bm l}$. The product $k_\alpha r^\alpha=r(1-\cos\theta)\simeq r\theta^2/2$
for small angles which is the case in the gravitational light-ray deflection experiments. Hence, equation (\ref{3}) tells us that the phase variation caused by space-time
difference $\delta x^\alpha=(c\delta t, \delta{\bm x})$ between two antennas is
\begin{equation}
\label{5}
\delta\phi=k_\alpha\delta x^\alpha+\frac{4GM_J}{c^2}\frac{\delta\theta}{\theta}\;,
\end{equation}
where $\delta\theta=-{\bm n}\cdot\delta{\bm x}/r$ with ${\bm n}={\bm
l}\times({\bm k}\times{\bm l})$ as the impact unit vector of the light
ray with respect to the retarded position of Jupiter ${\bm x}_J(s)$.
Currently undetectable terms of order $v_J/c$ have been neglected. The quantity $\delta
t$ is the measurable time delay and $\delta{\bm x}={\bm B}$
is a baseline between two VLBI stations. Since VLBI stations measure
the same wave front, $\delta\phi=0$. Thus, Eq. (\ref{5}) gives 
\begin{equation}
\label{6+}
\delta t=c^{-1}{\bm k}{\bm\cdot}{\bm B}+\Delta\;,
\end{equation}
where
\begin{equation}
\label{6}
\Delta =-\frac{4GM_J}{c^3r}\frac{{\bm n}{\bm\cdot}{\bm B}}{\theta}\;.
\end{equation} 
The post-Newtonian expansion of the retarded position of Jupiter, ${\bm x}_J(s)$, in Eq. (\ref{6}) with respect to the ratio $v_J/c$ (velocity of Jupiter/speed of gravity) around time of photon's arrival $t$ yields \cite{kop1,kop2} 
\begin{equation}
\label{q1}
\Delta =\Delta_S+\Delta_R\;.
\end{equation}
Here 
\begin{equation}
\label{q2}
\Delta_S=-\frac{4GM_J}{c^3}\frac{({\bm N}{\bm\cdot}{\bm B})}{\Theta}\;,
\end{equation}
is the static (instantaneous-like) part of the Shapiro time delay, and the gravity retardation $v_J/c$-correction
\begin{equation}
\label{7}
\Delta_R =\frac{4GM_J}{c^4R\Theta^2}\biggl[2
({\bm N}\cdot{\bm v}_J){\bm N}-{\bm k}\times({\bm v}_J\times{\bm k})\biggr]{\bm\cdot}{\bm B}\;,
\end{equation}
where ${\bm N}={\bm p}\times({\bm k}\times{\bm p})$ is the impact unit vector of the light ray with respect to the present position of Jupiter ${\bm x}_J(t)$ taken at the time of observation from JPL ephemeris \cite{standish}.
Eq. (\ref{7}) is the same as Eq. (4) from \cite{fk-apj}. It describes the post-Newtonian $v_J/c$ correction to $\Delta_S$ and can be detected because of the amplifying factor $\sim 1/\Theta^2$.   

Samuel \cite{samuel,sam2} incorrectly assumed that the experiment directly compared the
radio position of the quasar with that of Jupiter, and that the
direction of Jupiter was determined by light reflected from its
surface. This assumption would correspond to direct measurement of the
angle $\theta$ and hence no $v_J/c$ terms would be observed since they are not evident in Eq. (\ref{6}).  The
experiment, however, monitored the position of the quasar as a
function of the atomic time by the arrival of the quasar's photons at the radio
telescope, while the Jupiter's position was not directly observed \footnote{Jupiter has too large angular size in the sky compared to VLBI directional diagram.} but determined separately via a
precise JPL ephemeris \footnote{JPL ephemeris of Jupiter is constructed from radio-tracking of spacecrafts orbiting Jupiter. Hence, the experiment \cite{fk-apj} effectively compared the retarded position of Jupiter ${\bm x}_J(s)$ derived from the gravitational time delay with that obtained from the spacecraft's radio-tracking. We have used the same idea to measure the aberration of gravity in another VLBI experiment in October 2005 \cite{kop5}.} \cite{standish}, evaluated at the same atomic time as the arrival of
a radio photon from the quasar (via standard transformations from barycentric time to atomic
time \cite{bk,times}).  Hence the {\it actual} angle used in VLBI correlator for measuring $\Delta$ in equation (\ref{q1}) is $\Theta
$, not Samuel's $\theta_{obs}\equiv\theta$. Thus, the $v_J/c$ correction $\Delta_R$ was clearly separated in the VLBI correlator Calc/Solve program \cite{calc} from $\Delta_S$ and measured with a precision of 20\% \cite{fk-apj,kop1}. Additional critical analysis of the theoretical concepts involved to the interpretation of the Jovian light-ray deflection experiment can be found in \cite{kop4,kop3,kop6}. Their study can be useful for experts working on future projects like Gaia space mission \cite{gaia} and Square Kilometer Array \cite{ska}. 

We thank S. Samuel, G. Sch\"afer, S. Carlip, C.~M. Will for valuable comments and V.~B. Braginsky and M.~V. Sazhin for fruitful and constructive discussions.

\end{document}